%% file: cameraready.tex
\documentclass[conference]{IEEEtran}
\usepackage{cite}
\usepackage{amsmath,amssymb,amsfonts,mathtools}
\usepackage{algorithmic}
\usepackage{graphicx}
\usepackage{textcomp}
\usepackage{xcolor}

\def\BibTeX{{\rm B\kern-.05em{\sc i\kern-.025em b}\kern-.08em
    T\kern-.1667em\lower.7ex\hbox{E}\kern-.125emX}}
\renewcommand{\IEEEQED}{\IEEEQEDopen}

\newtheorem{theorem}{Theorem}

\newtheorem{remark}{Remark}
\begin{document}

\title{The Influence of Canyon Shadowing on Device-to-Device
  Connectivity in Urban Scenario
}

\author{\IEEEauthorblockN{Quentin Le Gall\IEEEauthorrefmark{1}, Bart\l{}omiej B\l{}aszczyszyn\IEEEauthorrefmark{2}, Elie Cali\IEEEauthorrefmark{1} and Taoufik En-Najjary\IEEEauthorrefmark{1}}
\IEEEauthorblockA{\IEEEauthorrefmark{1}Modelling and Statistical Analysis, Orange Labs Networks, Ch\^{a}tillon, France\\
Email: quentin1.legall@orange.com, elie.cali@orange.com and taoufik.ennajjary@orange.com}
\IEEEauthorblockA{\IEEEauthorrefmark{2}Inria-ENS, Paris, France
Email: bartek.blaszczyszyn@ens.fr}}

\maketitle

\begin{abstract}
In this work, we use percolation theory to study the  feasibility  of large-scale  connectivity  of  relay-augmented device-to-device (D2D)  networks  in an urban scenario featuring a haphazard system of streets and  canyon shadowing allowing only for line-of-sight (LOS) communications in a   finite range.
We use a homogeneous  Poisson-Voronoi  tessellation (PVT) model of streets with   homogeneous Poisson users (devices)  on its edges 
and independent Bernoulli relays on the vertices. 
Using this model, we demonstrate the existence of a minimal threshold for relays below which large-scale connectivity of the network is not possible, regardless of all other network parameters. Through simulations, we estimate this threshold to 71.3\%. Moreover,  if the mean street length is not larger than  some threshold (predicted to 74.3\% of the communication range; which might be  the case in a typical urban scenario) then any (whatever small) density  of  users  can  be  compensated by equipping more  crossroads  with relays.
Above this latter threshold,
good connectivity requires some minimal density of users, compensated by the relays
in a way we make explicit. The existence of the above regimes brings interesting qualitative  arguments to the discussion on the  possible D2D deployment  scenarios.
\end{abstract}


\section{Introduction}
The fifth generation (5G) of mobile networks currently concentrates an
intensive research effort covering broad fields such as security,
energy consumption, radio communications or resource allocation
\cite{andrews2014will, boccardi2014five,wang2014cellular}. One of the main technical challenges of 5G remains to
face the exponential growth of mobile data traffic while keeping up
with the quality of service (QoS). Device-to-Device (D2D) is deeply
investigated in this regard \cite{tehrani2014device}. Coverage extension could
also be achieved thanks to multihop D2D networks \cite{lin_comprehensive_2013}. This
opens the way to crowd networking and uberization of telecommunications networks \cite{asadi_survey_2014}, which represent high economic stakes. 

As a matter of fact, studying the technical feasibility of
large-scale connectivity of D2D networks seems critical for
operators. To this end, resorting to mathematical models amenable to
numerical simulations remains a safe and necessary prelude to massive
investments. Since the seminal paper~\cite{gilbert1961random} of Gilbert, the
question of large-scale connectivity in telecommunication networks has mathematically been dealt with using percolation theory \cite{meester_continuum_1996,grimmett2013percolation}.
Recent refinements have taken into account various street system models as the support of the network \cite{gloaguen2006fitting}. 

In this paper, as the main novelty of our work, we consider the {\em canyon effect
  of shadowing}  allowing  only for  line-of-sight (LOS) connections
on the streets: only network nodes located on
the  same street, and whose relative distance is less than a certain
threshold can establish communication. We then apply this assumption to an existing street model and study its impact on the connectivity properties of a D2D network using percolation theory. \medskip

\emph{The main results of this paper are the following ones:}

The canyon shadowing assumption, combined with our model for streets and network users, requires the presence of relays located at crossroads in order to achieve connectivity between adjacent streets. We prove that there exists a minimal fraction of crossroads which have to be equipped with relays below which 
large-scale connectivity of the D2D network cannot be achieved regardless of all
  other network parameters. 
\indent Regarding the interplay between the street system and the transmission range of D2D technology, we exhibit two different connectivity regimes for our model: one where large-scale connectivity can solely rely on relays, the other one where a high enough density of users, compensated by relays, is required.
\medskip

{\em The remaining part of this paper is organized as follows:} We begin in Section~\ref{s.RelatedWorks} by recalling related works. Next, in Section~\ref{s.Model}, we present our  model with its  associated assumptions. Then, in Section~\ref{s.Results}, we present our results.
Finally, we conclude our work in Section~\ref{s.Conclusions}.

\section{Related works}
\label{s.RelatedWorks}
In his founding work \cite{gilbert1961random}, Gilbert modelled a wireless network by a random graph and interpreted large-scale connectivity as percolation of this graph, i.e. the existence of an infinite connected component with positive probability.
However, this first model did not include any geometric features nor propagation effects.

The impact of fading and
interference in Gilbert's model was studied in \cite{dousse2005impact,dousse2006percolation}. In these works, the authors considered new connectivity conditions: a connection between two nodes of the network depends not only on their relative distance anymore, but also on the position of all other nodes of the network through the signal-to-interference plus noise ratio (SINR). 

The influence of the geometric features of the considered territories and simulation perspectives have been considered in \cite{gloaguen2006fitting, gloaguen2009parametric}. In these works, real street systems are fitted by random tessellations, including Poisson-Voronoi tessellations (PVT), more amenable to statistical analysis.

Connectivity for D2D networks on street systems using percolation
theory was explored only recently in~\cite{cali2018percolation}, building on the theoretical results from~\cite{hirsch_continuum_2017}
regarding percolation of Cox models.   
Very recently, percolation  of  the  SINR  graph  associated with Cox processes has been studied in~\cite{SINR-Cox2018}.
Cox processes 
cluster their points more than Poisson point processes~\cite{blaszczyszyn2014comparison}
and, in general, their percolation  properties cannot be simply derived by a comparison to this latter model~\cite{blaszczyszyn2013clustering}.
Our work is also related to~\cite{ziesche2016bernoulli} where Bernoulli percolation on random tessellations, including PVT, is studied.

Completely different self-similar  street systems with canyon shadowing effects have been considered in \cite{jacquet2017selfIEEE, jacquet2017selfSpringer}.

Regarding D2D per se, the surveys \cite{asadi_survey_2014, gandotra2016device} exhibit a rich variety of use cases for D2D communications.
Technical promises and contributions of D2D to 5G networks are investigated in \cite{tehrani2014device}. Many more questions regarding D2D deployment scenarios have been explored  \cite{fodor2012design,janis2009interference,yu2011resource, lin2014overview}. Technical issues related to D2D development are out of the scope of this paper.
\section{Network Model}
\label{s.Model}
We first present the crucial system assumptions and then describe our percolation model of large-scale connectivity for D2D communications.
\subsection{System assumptions}
\label{assumption}
Several assumptions have been made in our model, either for physical reasons or for the sake of simplicity.  \\
\indent Reflections of the waves on the buildings and the crossroads as well as diffractions on the edges of the buildings were not considered as a first step, for simplicity. Therefore, we modelled the street system as a two-dimensional tessellation. \\
\indent As in \cite{glauche_continuum_2003}, we assume a constant communication radius. This implies that we assume the transmission power of all devices and network relays to be constant and equal to a global common value. We also neglect any interference phenomenon or user mobility. 
The connectivity mechanism 
of our model only allows for LOS communications between a source and a target (whether they are a relay or an actual user equipped with a device): this is the canyon shadowing assumption. 
This implies that the physical obstacles encountered in our model are sufficiently absorbing to prevent any signal from being transmitted through them. In the context of 5G, where the main part of the useful spectrum consists of very high frequencies, this is indeed the case. 

\subsection{Description of the model}
\begin{figure}[t!]
\vspace{-7ex}
\centerline{\includegraphics[width=\linewidth]{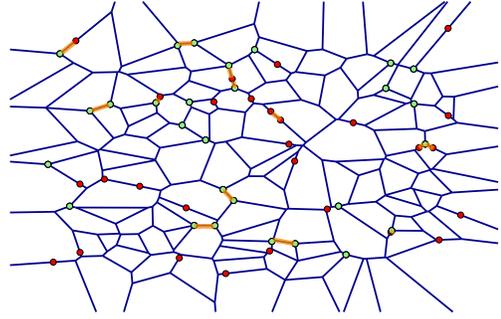}}
\vspace{-7ex}
\caption{Example of a D2D network simulated using our model. The blue lines represent the edges of the Poisson-Voronoi tessellation  modeling the street system. The red points represent the network users. 
The green points represent the relays. 
Possible connections are highlighted in orange: these correspond to pairs of points being in line-of-sight and of relative Euclidean distance smaller than some threshold~$r$.}
\label{Fig.network}
\vspace{-3ex}
\end{figure}
The network model relies on three major elements
 depicted on  Figure~\ref{Fig.network}
and detailed in what follows:
the street model, the respective distributions of users and relays and the D2D connectivity mechanism. 

\subsubsection{Street system}
\indent First, following 
\cite{gloaguen2006fitting}, we model the street system by a planar Poisson-Voronoi tessellation (PVT) $S$ generated by a homogeneous Poisson point process (PPP) in $\mathbb{R}^{2}$ of intensity $\lambda_{S} > 0$. 
We denote by $V$ (respectively $E$) the set of vertices (respectively edges) of $S$. Letting $\nu_{1}(S \cap B)$ be the total edge length of $S$ in any observation window defined by a Borel set $B$, we  denote by $\gamma \coloneqq \mathbb{E}\left[\nu_{1}\left(S \cap \left[-1/2;1/2\right]^{2}\right)\right]$ the total street length per unit area, expressed in km/km\textsuperscript{2}. It is  known that $\gamma = 2\sqrt{\lambda_{S}}$\,~\cite{okabe2009spatial}. Typical encountered values are $\gamma \approx 20 \: \text{km/km}^{2}$ for a city center of a classical European major city, while $\gamma \approx 1  \: \text{km/km}^{2}$ for rural areas.
Since $\gamma$ is an intrinsic characteristic determined by geographical location, we will consider it to be a fixed parameter of the problem considered here.  \medskip 
\subsubsection{Devices and relays distribution}
\indent Users are equipped with mobile devices and distributed according to a Cox point process $X$ driven by the random intensity measure $\lambda \nu_{1}(S \cap dx)$, where $\lambda \geq 0$ is the user intensity expressed in km\textsuperscript{-1} (the case $\lambda = 0$ corresponds to an absence of users and a D2D network only relying on relays placed by operators). Equivalently, whenever $\lambda > 0$, this means that conditioned on any realisation $S$ of the street system, $X$ is a Poisson point process with mean measure $\lambda \nu_{1}(S \cap dx)$. In particular, for any street segment $e \in E$, the number of users located on $e$ is a Poisson random variable with parameter $\lambda \nu_{1}(e)$ and all users are spread independently and uniformly on $e$. \\
\indent Network relays are placed on the crossroads of the street system according to a Bernoulli point process $Y$ of parameter $p \in \left[0,1\right]$. In other words, for each $v \in V$, a relay is placed at $v$ with probability $p$, independently from the state of any other crossroad in $V \setminus \lbrace v\rbrace$. \\
\indent The point process of users $X$ and the one of relays $Y$ are also assumed to be independent. \medskip

\subsubsection{Connectivity conditions}
\indent We assume a constant communication radius $r >0$ (expressed in kilometers) as in \cite{glauche_continuum_2003}. Letting $Z \coloneqq X \cup Y \coloneqq \lbrace Z_{i} \rbrace$ denote the superposition of the users and the relays point processes, two distinct network agents (either relays or users' devices) are connected by a D2D link if and only if they are in LOS and of relative Euclidean distance smaller than $r$, i.e. :
\begin{equation}
\label{connectivity_mechanism}
\forall \, i \neq j, \: Z_{i} \leftrightsquigarrow Z_{j} \Leftrightarrow 
\left\{
\begin{array}{l}
\exists \, e \in E, \, Z_{i} \in E  \  \text{and} \  Z_{j} \in E \\
\lVert Z_{i} - Z_{j} \rVert \leq r
\end{array}
\right.
\end{equation}
\indent The network is then represented by the connectivity graph 
whose vertices are the points of $Z$ and where an undirected edge $\lbrace Z_{i},Z_{j}\rbrace, \, i \neq j $ is drawn if and only if $Z_{i} \leftrightsquigarrow Z_{j}$. Connectivity of the network relying on the possibility of establishing long-range communications, we are thus interested in assessing whether there exists an infinite connected component of the connectivity graph for a given set of model parameters.
Some intrinsic scale-invariance  properties of our model allow us to reduce the number of these parameters, as presented in what follows. \medskip
\subsubsection{Dimensionless parameters of the model}
Similarly to \cite{cali2018percolation}, our model features scaling invariances: the Bernoulli process of relays is by definition motion-invariant \cite{chiu_stochastic_2013}, while changing $\gamma$ to $a\gamma$ for $a>0$ is equivalent to zooming or unzooming to a rescaled simulation window where $\lambda$ has changed to $a\lambda$ and $r$ to $r/a$. Therefore, the two dimensionless parameters $\lambda / \gamma$ and $r\gamma$ are scale-invariant. It is however physically more interesting to consider the following parameters, which are dimensionless and scale-invariant as well:
\begin{equation}
\label{dimensionless-parameters}
U = \frac{4}{3}\frac{\lambda}{\gamma} \quad \text{and} \quad H=\frac{4}{3}\frac{1}{r\gamma}    
\end{equation}
Indeed, following \cite[Section 9.4]{chiu_stochastic_2013}, $U$ represents the mean number of users per typical edge of the PVT street system, while $H$ is the mean number of hops necessary to ensure connectivity of a typical edge of the PVT street system. Note that $U$ both depends on the density of the street system and the density of users, while $H$ represents the interplay between the street system and the transmission range related to D2D technology.
The connectivity graph representing the D2D network will be denoted by $\mathcal{G}_{p,U,H}$.  
\subsection{Simulation method}
All of our numerical experiments have been performed using the statistical software R. Since an infinite graph cannot be simulated, we chose a squared simulation window of side $win$, expressed in kilometers. When possible, $win = 30 \: \text{km}$, a value chosen sufficiently large in practice so as to avoid any effects due to the finiteness of the simulation window. We chose not to simulate on a torus-traced window. Indeed, in \cite{cali2018percolation,mertens_continuum_2012}, the authors showed that simulations on a torus-traced window take much longer time for a very small gain in precision. \\
\indent For a set of parameters $(p,U,H)$, we first simulate a PVT $S$ with the desired parameter $\gamma$. Then, we label each street segment with a unique number. Thereafter, we simulate the corresponding users' Cox point process
and the relays' Bernoulli process (recall that $\lambda$ is determined by \eqref{dimensionless-parameters} and $p$ is given).
Each user is located on a unique street, while each relay is located on a crossroad at the intersection of 3 streets almost surely, see \cite{okabe2009spatial}. We assign to each user (respectively each relay) the label of the unique street (respectively the label of the streets) it is located on. As a matter of fact, two network agents are in LOS if and only if they share a common label. Determining all existing connections in an optimized way is thereafter straightforward: arrange the simulated network agents by street segment label, only keep the street segments containing at least two distinct network agents and then, for each street segment, compute the successive distances from one agent to the next one. If only one disconnection 
occurs on a given street segment, then it is not necessary to continue the computations for this street segment. We then keep track of the connected components of the simulated graph by using a union-find algorithm, as suggested in \cite{newman2001fast}. Finally, we declare that the simulated connectivity graph $\mathcal{G}_{p,U,H}$ percolates if there exists a left-right or a top-bottom crossing of the simulation window by a connected component. We then repeat this process 100 times (simulations showed that a greater amount of times does not enhance the precision of the results significantly) and are thus able to compute the proportion of simulations where the graph $\mathcal{G}_{p,U,H}$ percolates for a given set of parameters $(p,U,H)$.  \medskip
\indent 


\section{Results}
\label{s.Results}
We now present theoretical and numerical results of the study of our model.

\subsection{Minimal relay proportion}
\label{relay-influence}

\begin{figure}[t!]
\vspace{1ex}
\centerline{
\includegraphics[width=0.48\linewidth]{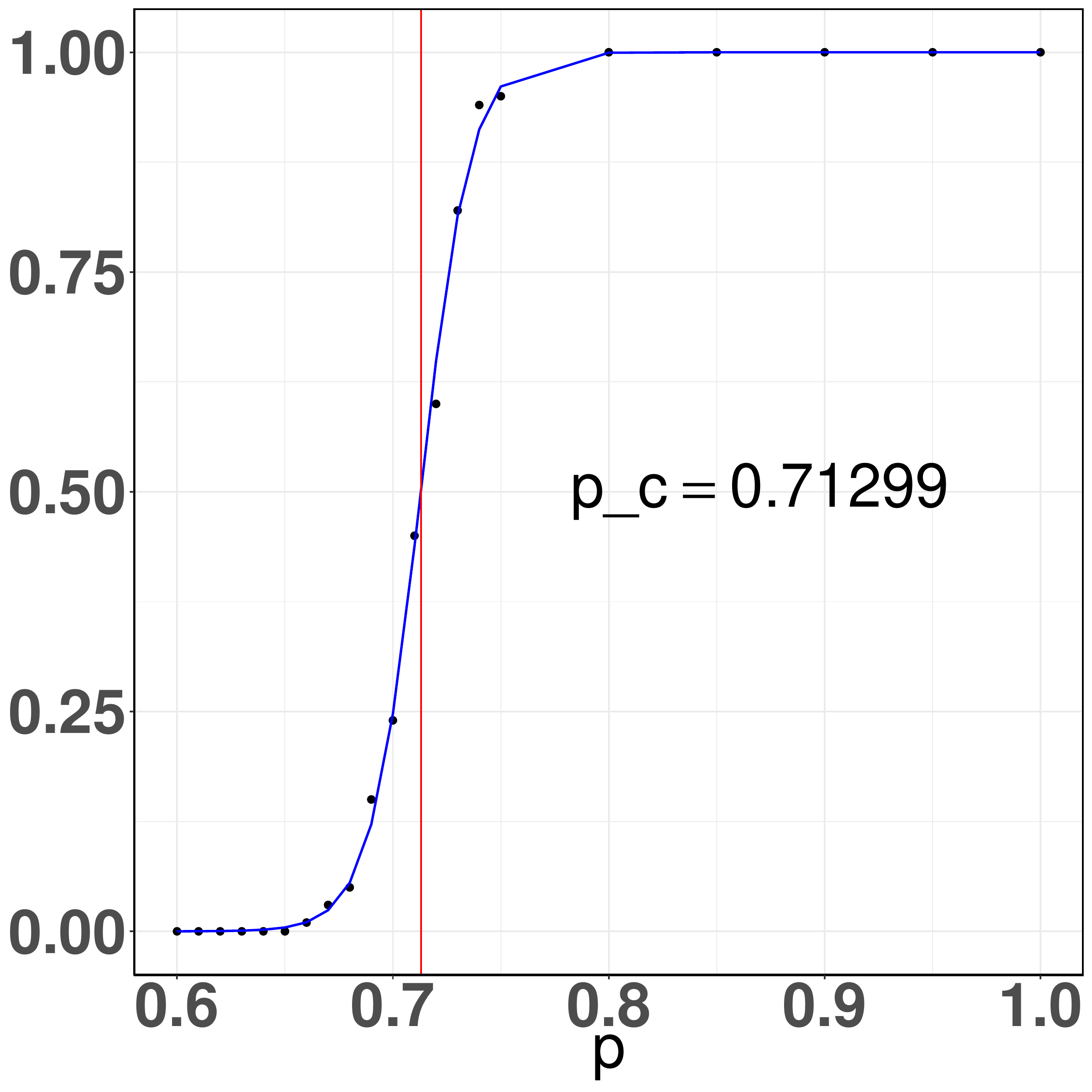}
\hspace{0.01\linewidth}
\includegraphics[width=0.48\linewidth]{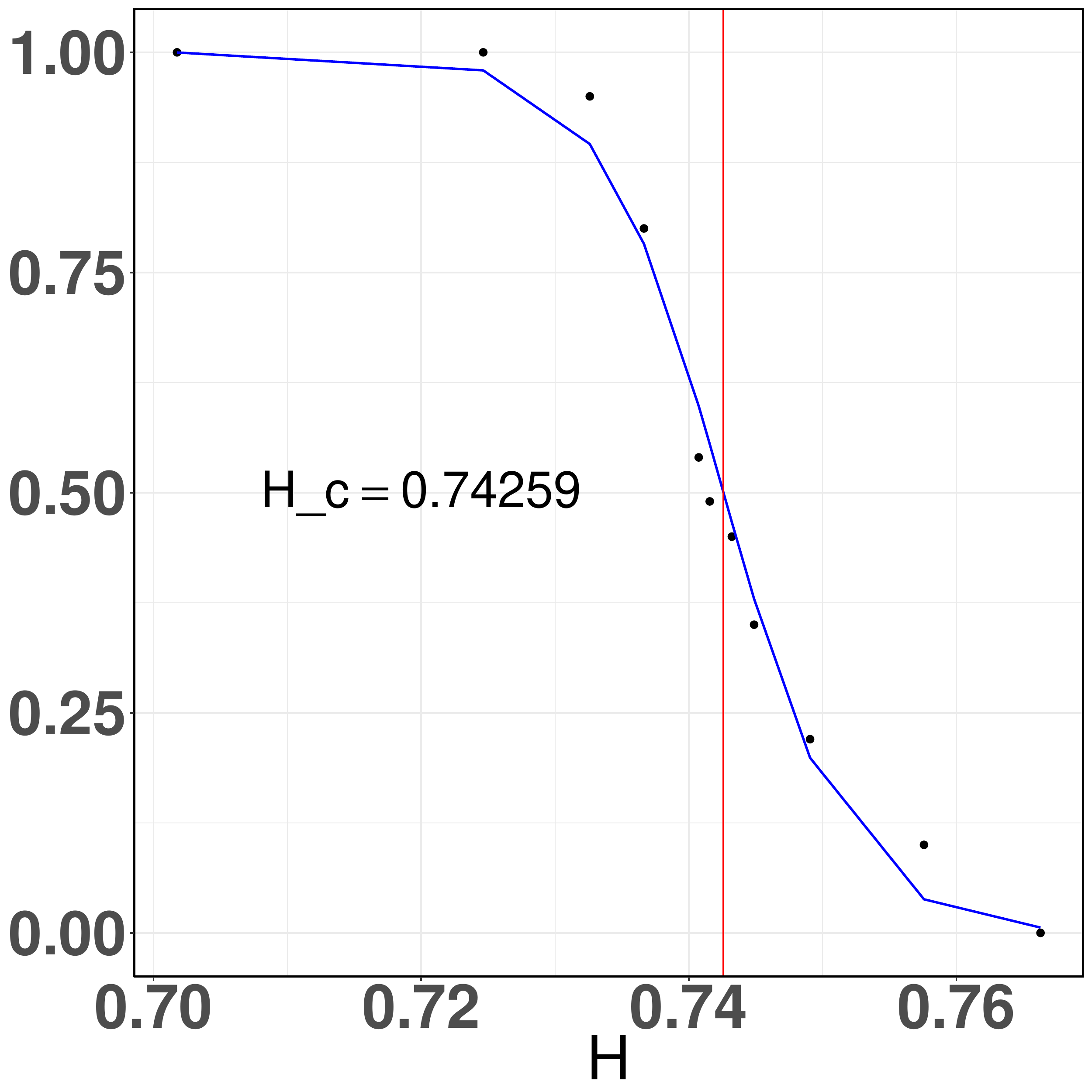}}
\vspace{-1.25ex}
\centerline{\footnotesize\hspace{0.25\linewidth} (a)\hspace{0.5\linewidth} (b) \hspace{0.25\linewidth}\ }
\vspace{-1.25ex}
\caption{Left: Estimation of $p_c$. Right: Estimation of $H_c$.
The points are the discrete values of the window-crossing probability
crossing obtained by simulation (window size 30x30 $\text{km}^{2}$, street density $\gamma = 20 \; \text{km/km}^{2}$), the curve is the logistic model and the vertical line is supposed to intercept the percolation threshold.}
\label{p-H-thresholds}
\vspace{-3ex}
\end{figure}

\indent Our first theoretical result is a minimality condition on  $p$ for the possibility of percolation of the connectivity graph $\mathcal{G}_{p,U,H}$. 
To this end, we consider another percolation model: the Bernoulli site percolation model~
on $S$. In this model, each vertex of $S$ is either open (i.e. present) with probability $p \in \left[0,1\right]$ or closed (i.e. absent) with probability $1 - p$. 
Denote by $\tilde{\mathcal{G}}_{{p}}$ the subgraph of $S$ obtained by only keeping the open vertices of $S$ (i.e. $\lbrace v \in V: v \; \text{is open} \rbrace$) and the edges of $E$ connecting them. As usual, say that $\tilde{\mathcal{G}}_{{p}}$ percolates 
if it has an infinite connected component. Define as usual the percolation threshold:
\begin{equation}
\label{critical_site}
p_c \coloneqq p_{c}^{\text{site, PVT}} \coloneqq \inf \lbrace {p} \geq 0, \; \tilde{\mathcal{G}}_{p} \text{\: percolates} \rbrace
\end{equation}
It is known that 
$p_c$ is independent of $\lambda_{S}$ and that $p_c \in (0,1)$
\cite{ziesche2016bernoulli}.
Moreover, \cite{neher2008topological} found the following theoretical estimate: $p_{c}^{\text{site, PVT}} \approx 0.7151 $, while \cite{becker_percolation_2009}, using Monte-Carlo simulations with periodic boundary conditions, numerically determines $p_{c}^{\text{site, PVT}} \approx 0.71410 \pm 0.00002$. We also performed Monte-Carlo simulations on our own to check the precision of our simulations. The results are shown in Figure~\ref{p-H-thresholds}(a).
A logistic model\footnote{Logistic regression consists in  estimating 
parameters $a$ and $b$ such that 
\begin{equation*}
\log\left (\frac{f(p)}{1-f(p)}\right)=ap+b, 
\end{equation*} where discrete values of $f(p)$ are obtained by simulations.} 
seems to fit a good approximation.
Since the theoretical curve on an infinite tessellation would be a 0-1 curve with cutoff value $p_c$ by ergodicity, we can reasonably approximate $p_c$ by the abscissa of the inflection point of the logistic curve, yielding:   
\begin{equation}
\label{e.pc}
p_c:=p_{c}^{\text{site,PVT}} \approx 0.71299 \;  
\end{equation}
This is a fairly reasonable approximation for our purposes.

\indent 
By comparing percolation of the connectivity graph $\mathcal{G}_{p,U,H}$ with Bernoulli site percolation on $S$, we obtained the following result: \medskip
\begin{theorem}[Minimality condition on $p$] \label{minimality_theorem} If 
$p<p_c$, then, for all $U \geq 0$ and $H>0$, the connectivity graph $\mathcal{G}_{p,U,H}$ does not percolate, i.e. long-distance multihop D2D communications are not possible.
\end{theorem}
\begin{IEEEproof} 
Let $p < p_c$.
Consider site percolation on $S$ with parameter $p$. Then, by \eqref{critical_site} the associated graph $\tilde{\mathcal{G}}_{p}$ does not percolate. But since $\mathcal{G}_{p,U,H}$ is a subgraph of $\tilde{\mathcal{G}}_{p}$ for all $U \geq 0$ and $H>0$, the absence of percolation of $\tilde{\mathcal{G}}_{p}$ implies the absence of percolation of $\mathcal{G}_{p,U,H}$. Hence the result.
\end{IEEEproof}
\vspace{\baselineskip}

\begin{remark}
Theorem~\ref{minimality_theorem} has the following practical consequence: an operator willing to constitute a multihop D2D network should equip an important number of crossroads. 
This represents a heavy investment, which needs to be counterbalanced. Only relying on users' devices to allow for long-distance connectivity is not a viable option. Finally, note that our result ensures a minimality condition on $p$ only. Indeed, we shall see in what follows that 
there exists a regime of network parameters, such that even when $p=1$, i.e. all crossroads are equipped with relays, the connectivity graph does not percolate in the absence of users. A matching maximality result on $p$ is therefore unthinkable.
\end{remark}

\subsection{Relay-limited connectivity}
\label{ss.RelayLimited}
After having proven that there exists a minimal relay proportion under which no large-scale connectivity of the network is possible regardless of all other network parameters, we may wonder whether connectivity of the D2D network can solely rely on relays. Indeed, with  the D2D communication range being a physical constraint imposed by the type of D2D technology, in  particular the type of the radio link \cite{asadi_survey_2014}, one can think  that if there are sufficiently many streets shorter than this range, a sufficiently high proportion of relays could  be deployed, allowing for long-range D2D communications even when the user density is low, i.e. $U \rightarrow 0$. 
This is indeed the case, as we shall see in what follows.

\indent For given $H$,
let us consider first  the best possible case where all crossroads are equipped with relays, i.e. $p=1$. If large-scale connectivity without users cannot be achieved when all crossroads are equipped with relays, then it also cannot be achieved for any $p \in (p_c,1)$. Setting $p=1$, define the following critical value for $H$: 
\begin{equation}
    \label{critical-H}
    H_{c}  \coloneqq \sup \lbrace H > 0, \; \mathcal{G}_{1,0,H} \; \text{percolates}\}\,.
\end{equation}
Checking the possibility of percolation in the absence of users is equivalent to 
verifying whether $H_c>0$. This theoretical question can be
answered affirmatively using theoretical tools out of the scope of this paper, and we approximate $H_c$ by simulations.

In this regard, we compute, for a grid of values of $H$, the proportion $g(H)$ of simulations where the graph $\mathcal{G}_{1,0,H}$ percolates. Here, an inverse sigmoid yields a good fitting of the estimated curve, see Figure~\ref{p-H-thresholds}(b). Finally, we recover $H_{c}$ by the abscissa of the inflection point of the logistic curve and find the following estimate: $H_c \approx 0.743$.   \medskip

\indent In the remaining part of this section, we investigate the relay-limited connectivity regime 
($H < H_c$), that is when there is a possibility of percolation in the absence of users. The question is whether we need the complete deployment of  
relays ($p=1$) for percolation, as assumed in~\eqref{critical-H}.
The intuition is that statistically shorter streets (corresponding to $H<H_c$) might require only some proportion of crossroads equipped with relays. In mathematical terms, we define the following critical proportion of relays ensuring percolation of the connectivity graph $\mathcal{G}_{p,0,H}$ in the absence of users in the relay-limited connectivity regime  $H<H_c$:
\begin{equation}
    \label{p_c(H)}
    p_c(H) \coloneqq \inf \lbrace p \in (0,1), \; \mathcal{G}_{p,0,H} \text{\: percolates} \rbrace
\end{equation}
We already know from Section~\ref{relay-influence} that 
$p_c(H)\ge p_c$ and it can be proved mathematically that $p_c(H)<1$ for all $H<H_c$. Our goal is again to approximate this function by simulation.

The methodology is quite the same as in the previous numerical simulations: this time, for a given $H<H_c$ and a grid of values of $p \in (p_c,1)$, we compute the proportion $k(p)$ of simulations where the graph $\mathcal{G}_{p,0,H}$ percolates. Theory tells us that $k$ is increasing in $p$ (more relays indeed implies more connections, hence making percolation easier to occur) and the logistic model yields again a good fitting of the estimated curve, leading to $p_c(H)$ as the inflection point of the logistic curve. The estimated values are presented in Table~\ref{tab-p_c(H)}. As can easily be guessed and as is confirmed by our results, $p_c(H)$ is increasing with $H$.
We were only able to consider $H>0.46$. Below this value 
the system started having an erratic behaviour not giving any  
reasonable estimation of $p_{c}(H)$. This can be explained by the  fact
that when $H$ approaches~0, $p$ approaches $p_{c}$ and the simulation of the model close to criticality is much trickier.

\begin{remark}
In practice, operators have leverage on $p$ (by equipping more or less crossroads with relays). The results provided in Table~\ref{tab-p_c(H)}
allow them to find an appropriate proportion of relays
in function of  $H$, which depends both on the D2D technology and the inner geometry of the network. In this table we also relate $H$ to the D2D communication range $r=\frac{4}{3}\frac{1}{H\gamma}$  (see \eqref{dimensionless-parameters}) in case of an urban environment by taking $\gamma = 20 \, \text{km/km}^{2}$. 
It is worth noticing that the  relay-limited connectivity regime 
($H < H_c$)  in such an environment implies  $r$ smaller  than 150 meters in most cases. This is a technological threshold which does not seem physically unreachable \cite{lin_comprehensive_2013}. 
\end{remark}
\begin{table}[t!]
\vspace{0.5ex}
\caption{Critical parameter $p_{c}(H)$ and corresponding $r$ in an urban environment as a function of $H$}
\begin{center}
\begin{tabular}{|c|c|c|}
\hline
$H$ & $p_c(H)$ & $r$ (meters), urban environment \\
\hline
0.467 & 0.75 & 142.96  \\
0.487 & 0.76 & 136.96 \\
0.503 & 0.77 & 132.60 \\
0.521 & 0.78 & 127.95  \\
0.534 & 0.79 & 124.90 \\
0.548 & 0.80 & 121.72  \\
0.609 & 0.85 & 109.52 \\
0.655 & 0.90 & 101.75 \\
0.702 & 0.95  & 95.03 \\
$H_c \approx 0.743$ & 1 & 89.78  \\
\hline
\end{tabular}
\label{tab-p_c(H)}
\end{center}
\vspace{-4ex}
\end{table}
\begin{figure}[t!]
\centering
\includegraphics[width=0.50\linewidth]{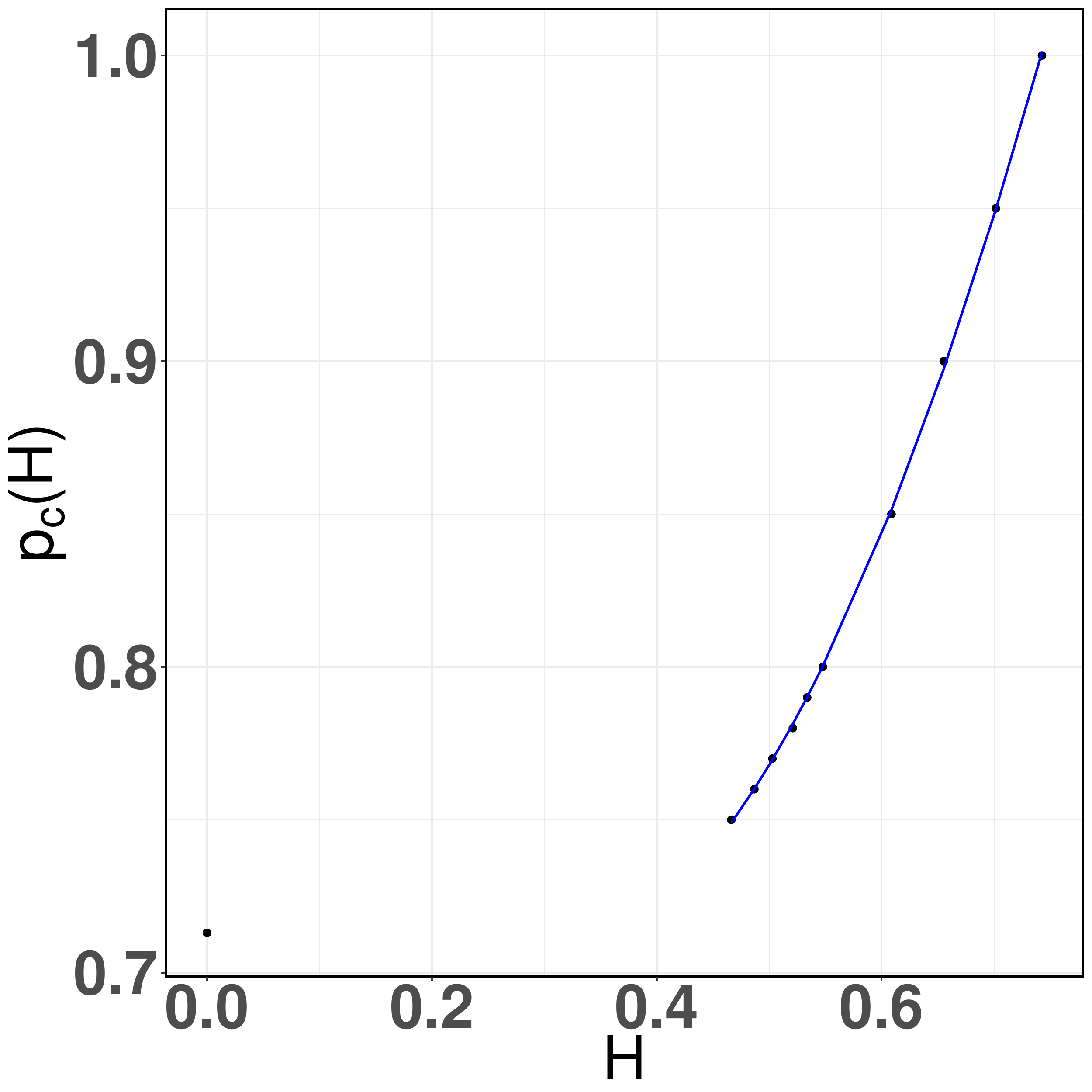}
\vspace{-1ex}
\caption{Critical relay proportion $p_c(H)$ in the absence of users ($U=0$) as a function of $H$. The  points are the discrete values from Table~\ref{tab-p_c(H)}, the curve is the estimated quadratic fit.
The isolated point $p_c(0)=p_c$ corresponds to the (absolute) minimal relay proportion~\eqref{e.pc}. }
\label{quadratic-fit-p_c(H)}
\vspace{-3ex}
\end{figure}

\indent  In order to get a continuous approximation of $p_c(H)$ for $0.46<H<H_c$ we interpolate the discrete values given in Table~\ref{tab-p_c(H)}. 
We found out that the following quadratic model 
$p_c(H) \approx aH^{2} + bH + c$
with the  value of the coefficients $a$, $b$ and $c$ estimated by linear regression
$a \approx 1.45$, $b \approx -0.84$, $c \approx 0.83$
yields a good fit able to explain 99\% of the variance. Fig.~\ref{quadratic-fit-p_c(H)} illustrates the discrete simulated curve and the estimated quadratic fit for $p_c(H)$, confirming that a quadratic model is a very good approximation when  $0.46<H<H_c$. We do not have any approximation of $p_c(H)$ for smaller $H$. However, we believe that it drops quickly to the (absolute) minimal relay proportion
$p_c(0)=p_c\approx 0.71299$ given by~\eqref{e.pc}.

\subsection{Relay-and-user-limited connectivity}
\label{ss.Relay-and-Users}
The main question arising from the previous section is about what happens when the D2D range and the street system do not allow to reach the critical parameter $H_c$ and thus to solely rely on relays for ensuring large-scale connectivity of the network. In other words, for $H>H_c$, is there a critical user density above which long-range communications are possible? If so, which minimal relay proportion is appropriate for ensuring large-scale connectivity with the help of users serving as D2D relays? \\
\indent As in Section~\ref{ss.RelayLimited}, for some $H>H_c$, let us consider first the case where $p=1$. If large-scale connectivity relying on both users and relays cannot be achieved when all crossroads are equipped with relays, then it also cannot be achieved for any $p \in (p_c,1)$. Setting $p=1$ and for given $H>H_c$, define the following critical value for $U$:
\begin{equation}
\label{critical-U}
U_c(H) \coloneqq \inf \lbrace U \geq 0, \; \mathcal{G}_{1,U,H} \; \text{percolates} \rbrace  
\end{equation}

\subsubsection{Non-triviality of the critical parameter $U_c(H)$}
On a theoretical perspective, we were able to prove that under sufficiently general conditions, the critical parameter $U_c(H)$ representing the minimal average number of users per typical street allowing for long-range communications is indeed positive and finite. Our result is the following one:
\\
\begin{theorem}[Non-triviality of $U_c(H)$]\label{critical-U-thm}
There exists a critical value $H^{*} \geq H_c$ such that whenever $H > H^{*}$ we have   \hbox{$0<U_c(H)<\infty$}. 
\end{theorem}
\medskip
Theorem~\ref{critical-U-thm} 
says that if the streets of the network are long enough compared to the D2D range ($H>H^{*}$),
then long-range communications can only be achieved under a sufficiently high (but finite) user density.
There is a possible theoretical gap between $H^*\ge H_c$ and the critical value $H_c\approx 0.743$ found in Section~\ref{ss.RelayLimited}, however our simulations suggest that $H^*\approx H_c$.
A rigorous proof of the above result 
follows the approach developed in~\cite{hirsch_continuum_2017}.
As the goal of this paper is more about giving numerical estimates, and due to space constraints, we only give a rough sketch of the proof.
\medskip
\begin{IEEEproof}[Sketch of proof of Theorem~\ref{critical-U-thm}]
The main problem faced in the study of percolation in a random environment (PVT street system in our case) is the spatial dependence of the environment.
By the  {\em stabilization property}~\cite{hirsch_continuum_2017} of the PVT, the configuration of the network environment in a given observation window only depends on a bounded region including the observation window with high probability. In other words, if two observed regions of the network are distant enough, they are independent. This allows one to introduce a discretized site percolation process featuring short-range spatial dependencies only. Well-chosen definitions of open and closed sites in the former process allow to ensure that if the discretized process does not percolate, then neither does $\mathcal{G}_{1,U,H}$. Finally using the domination by product measures theorem~\cite{liggett1997domination} allows one to conclude that the discretized process does not percolate if $U$ is sufficiently small and $H > H^{*}$ for some absolute constant $H^{*}\geq H_c$ ($H^{*}$ only depends on the edge length distribution of the edges in a PVT), thus proving that $U_c(H)>0$.

Similar techniques are used in the proof of the finiteness of $U_{c}(H)$.
In this case, we introduce a discrete percolation process chosen so that if the former process percolates, then so does $\mathcal{G}_{1,U,H}$. Crucially relying on the \emph{asymptotic essential connectedness}~\cite{hirsch_continuum_2017} of the PVT street system~$S$ and using again the domination by product measures theorem \cite{liggett1997domination} allows one to conclude that the discretized percolation process percolates if $U$ and $H$ are sufficiently large. Hence the result.
\end{IEEEproof}

\vspace{1\baselineskip}

\subsubsection{Numerical estimations of $U_c(H)$}
\begin{figure}[t!]
\vspace{0.5ex}
\centerline{
\includegraphics[width=0.47\linewidth]{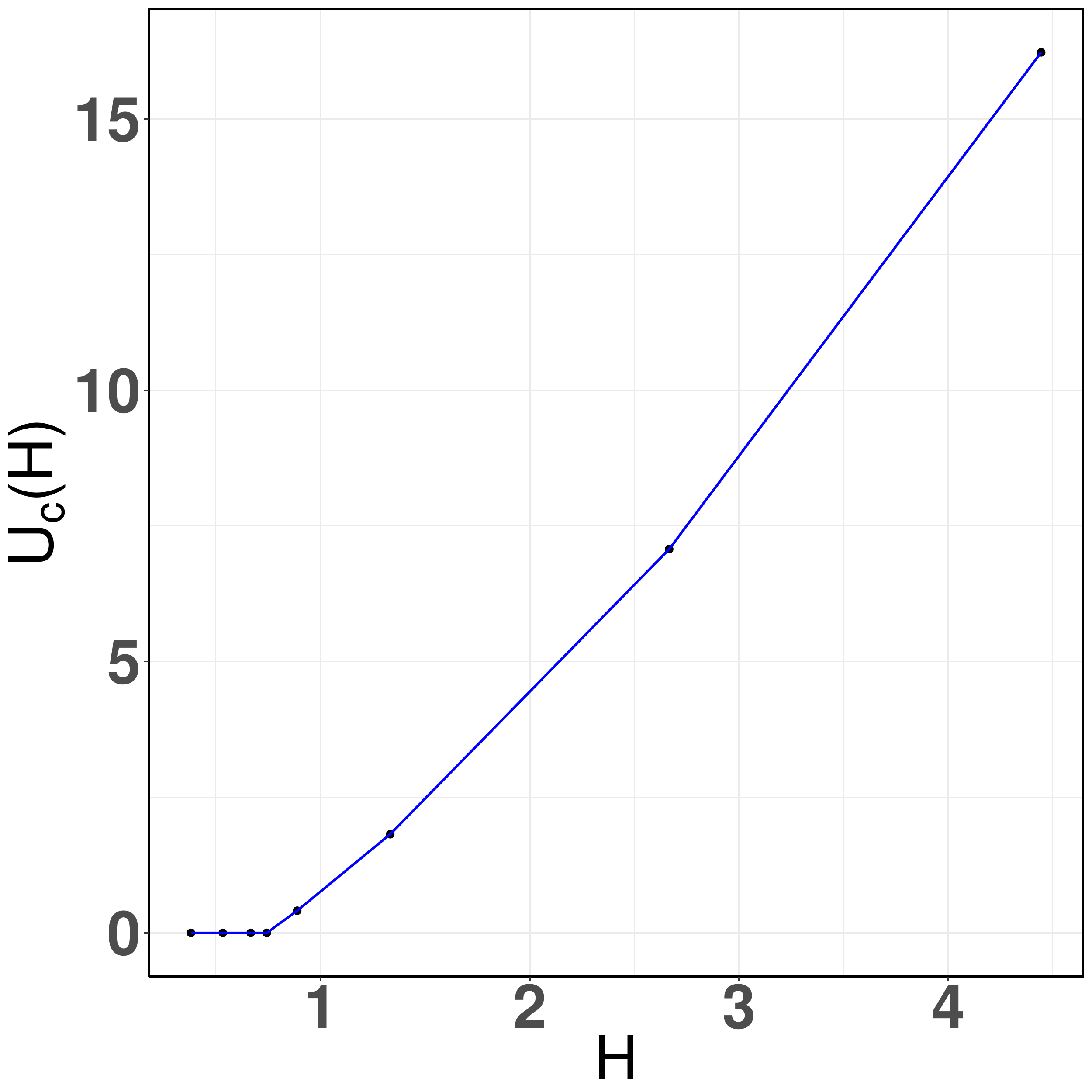}
\hspace{0.01\linewidth}
\includegraphics[width=0.47\linewidth]{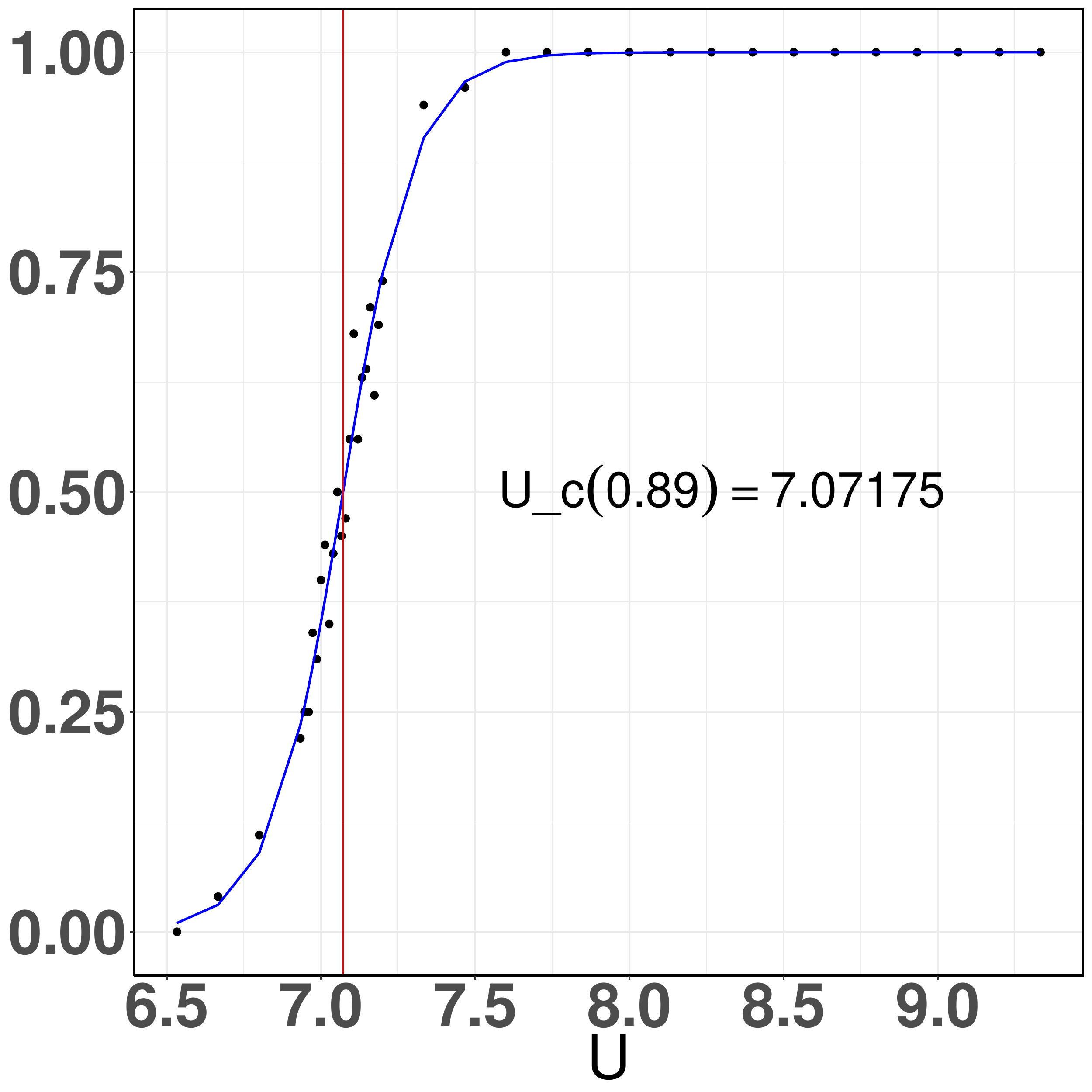}}
\vspace{-1ex}
\centerline{\footnotesize\hspace{0.25\linewidth} (a)\hspace{0.5\linewidth} (b) \hspace{0.25\linewidth}\ }
\vspace{-1ex}
\caption{Left: $U_c(H)$ as a function of $H$. Right: Example of estimation of $U_{c}(H)$ for $H\approx 0.89$. The simulation window is of size 10x10 $\text{km}^{2}$.
The points are the discrete values of the window-crossing probability obtained by simulations, the curve is the logistic model and the vertical line determines the intercept $U_{c}(H)$. }
\label{critical-user-estimation}
\vspace{-3ex}
\end{figure}

We now estimate the critical values $U_c(H)$ theoretically predicted in Theorem~\ref{critical-U-thm}.
The simulation method used to estimate $U_{c}(H)$ for given $H$ is merely the same as in Sections \ref{relay-influence} and \ref{ss.RelayLimited}: for a given $H$ and a grid of values for $U$, we simulate a large number of connectivity graphs $\mathcal{G}_{1,U,H}$ and compute the proportion $l(U)$ of simulations where $\mathcal{G}_{1,U,H}$ percolates. Again, a logistic model gives a good fitting of the estimated curve, and we determine $U_{c}$ by noting the abscissa of the inflection point of the logistic curve. Fig.~\ref{critical-user-estimation}(b) illustrates the estimation of $U_c(H)$ for $H \approx 0.89$ (corresponding to a D2D range $r = 75 \text{m}$). Fig.~\ref{critical-user-estimation}(a) provides such estimated values of $U_c(H)$ as a function of $H$.
Note that $U_c(H) = 0$ whenever $H < H_c \approx 0.743$. 

\subsection{Critical user density  in relay augmented D2D network}
\label{ss.NumericalEstimationsU}
From the results in Sections~\ref{ss.RelayLimited} and \ref{ss.Relay-and-Users}, we have seen that users and relays have to compensate each other to allow for arbitrarily long-range communications on the network whenever $H>H_c$. In fact, even when large-scale connectivity can solely be ensured by relays, i.e. $H<H_c$, an operator might rather want to invest less in relays and incentivize users to serve as D2D relays. The compromise between relay proportion and user density can be captured by either of the following functions:
\begin{itemize}
    \item $(U,H) \mapsto p_c(U,H) \coloneqq \inf \lbrace p>p_c, \; \mathcal{G}_{p,U,H}  \text{\: percolates} \rbrace$
    \item $(p,H) \mapsto U_c(p,H) \coloneqq \inf \lbrace U \geq 0, \; \mathcal{G}_{p,U,H}  \text{\: percolates} \rbrace$
\end{itemize}
Both approaches are actually equivalent and choosing either one is just a matter of convenience and practicality for numerical simulations. Indeed, it can mathematically be proven that $U_c(p,H)$ is a decreasing function of $p$ for fixed $H$ and can therefore be inverted: this leads back to the critical relay proportion $p_c(U,H)$.

\smallskip
\begin{remark}
\label{remark-inverse-Uc(p,H)}
The function $(p,H) \mapsto U_c(p,H)$ 
can be seen by an operator as an indicator of the average number of users needed to successfully deploy a D2D network for a given investment in relays.
Note that the function $U_c(H)$ defined in \eqref{critical-U} is such that for all $H>H_c, \, U_c(H) = U_c(p=1,H)$. The interest of computing $U_c(p,H)$ also when $H<H_c$ relies on the fact that an operator might rather want to rely on its already existing subscribers than on new relays.
\end{remark}
In what follows, we shall present   some values of $U_c(p,H)$ for both regimes $H<H_c$ and $H>H_c$.

\begin{figure}[t!]
\vspace{0.5ex}
\centerline{
\includegraphics[width=0.47\linewidth]{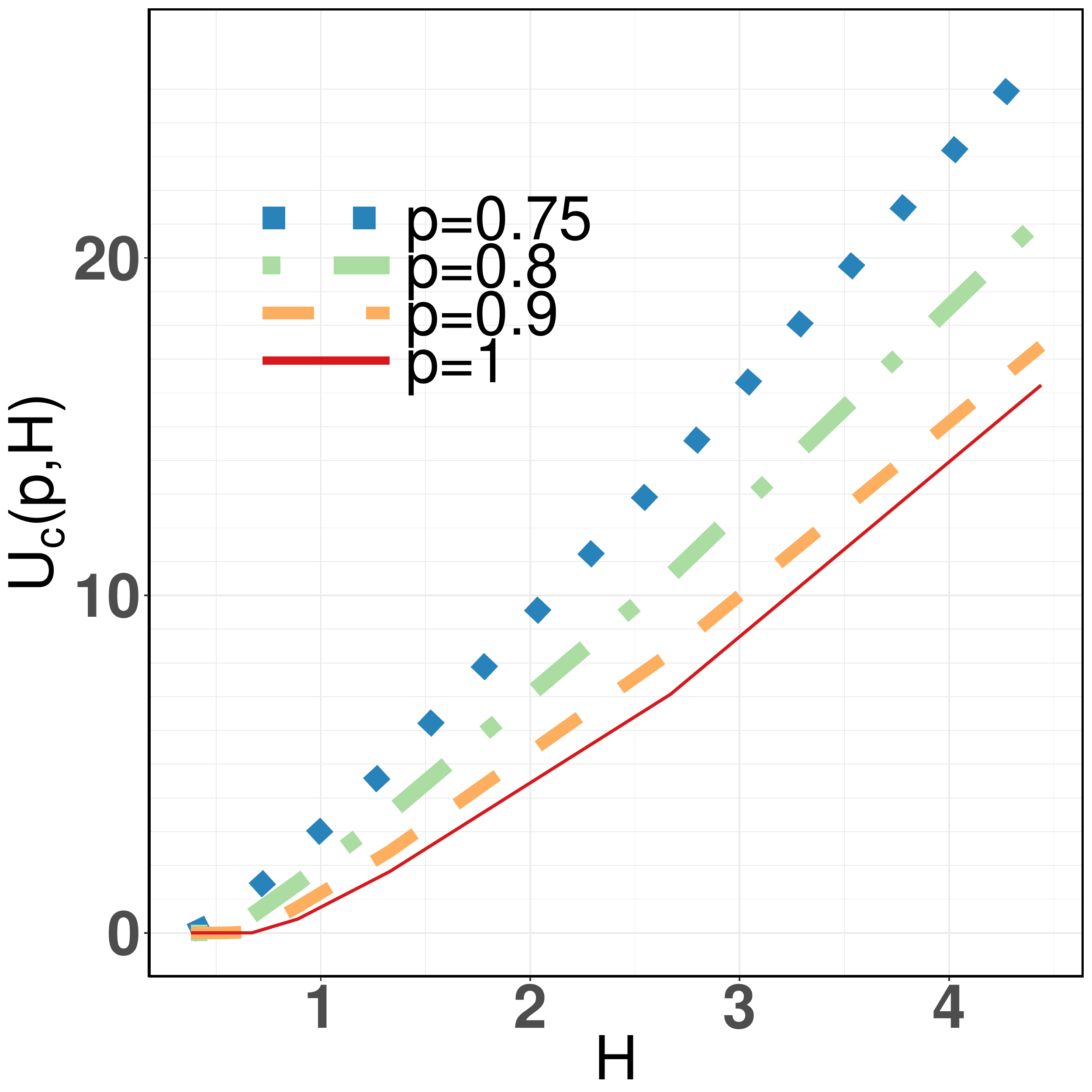}
\hspace{0.01\linewidth}
\includegraphics[width=0.47\linewidth]{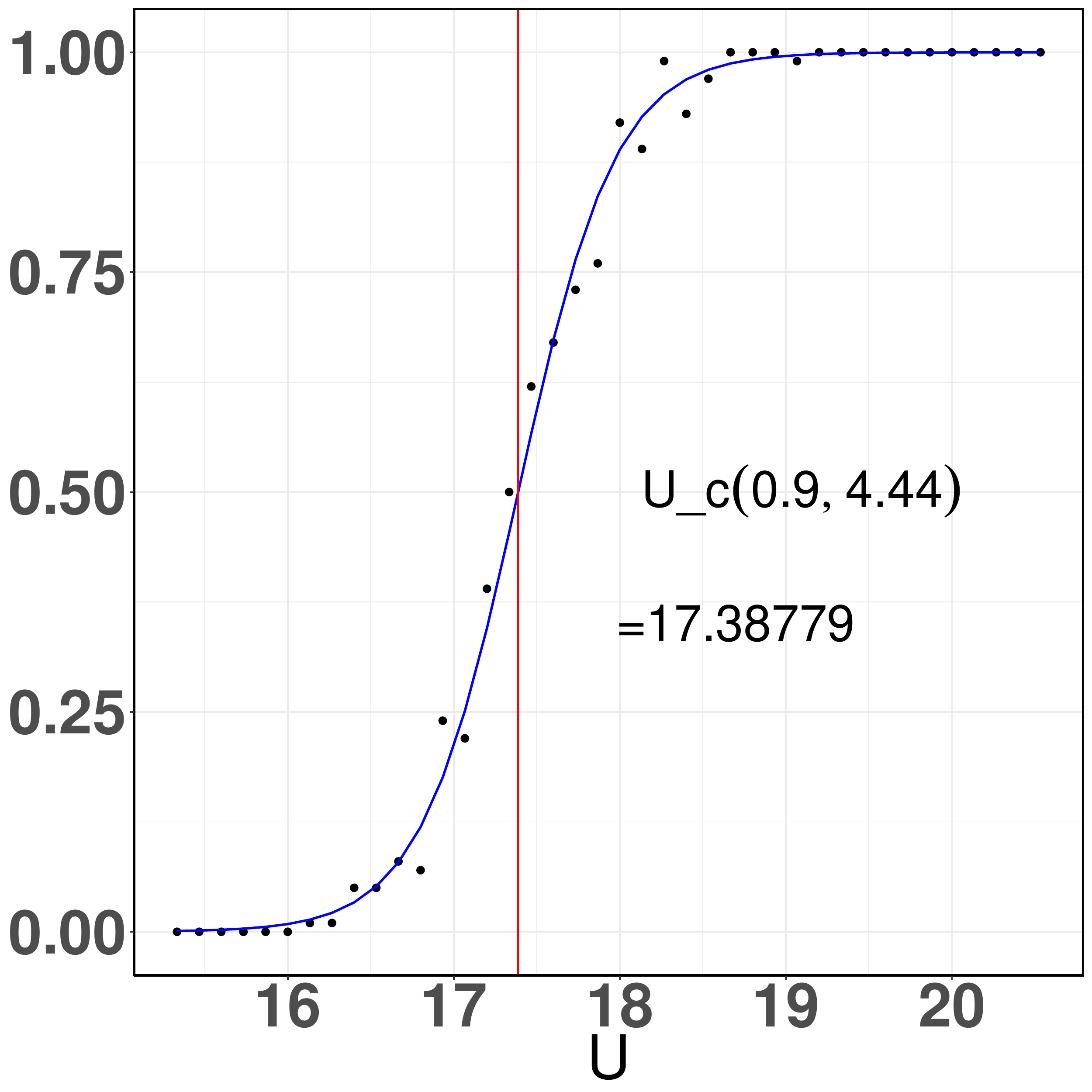}}
\vspace{-1ex}
\centerline{\footnotesize\hspace{0.25\linewidth} (a)\hspace{0.5\linewidth} (b) \hspace{0.25\linewidth}\ }
\vspace{-1ex}
\caption{Left: Critical user density $U_c(p,H)$ as a function of $H$ for several values of $p$. Right: Example of estimation of $U_{c}(p=0.9,H \approx 4.44)$, corresponding to $r=~15 \, \text{m}$ in an urban environment ($\gamma = 20 \, \text{km/km}^{2}$). The simulation window is of size 10x10 $\text{km}^{2}$. The points are the discrete values of the window-crossing probability obtained by simulations,  the curve is the logistic model and the vertical line determines the intercept $U_{c}(p,H)$. }
\label{critical-user-estimation-twovar}
\vspace{-2ex}
\end{figure}

\indent The simulation method used to estimate the critical average number of users $U_{c}(p,H)$ is merely the same as in Section~\ref{ss.Relay-and-Users}: for given $p>p_c$ and $H$, and for a grid of values for $U$, we simulate a large number of connectivity graphs $\mathcal{G}_{p,U,H}$ and compute the proportion $m(U)$ of simulations where $\mathcal{G}_{p,U,H}$ percolates. Again, a logistic model gives a good fitting of the estimated curve, and we determine $U_{c}(p,H)$ by noting the abscissa of the inflection point of the logistic curve. Fig.~\ref{critical-user-estimation-twovar}(b) illustrates an example. Results for estimations of $U_c(p,H)$ are given in Table \ref{tab-critical-U}. We also include a comparison with results from \cite{cali2018percolation}, where the authors simulated a model similar to ours without any shadowing effects (NoSha), i.e. there are only users on streets distributed according to a Cox process and any two users (being in LOS or not) with reciprocal Euclidean distance less than $r$ are connected. It is clear from Table \ref{tab-critical-U} that the previous estimates from \cite{cali2018percolation} are much smaller than ours: taking the canyon shadowing assumption into account in our model indeed provides more realistic information for operators.


\begin{table}[t!]
\caption{Critical user density $U_{c}(p,H)$ as a function of $p$ and $H$.}
\begin{center}
\vspace{-1ex}
\begin{tabular}{|c|c|c|c|c|c|}
\hline
\multicolumn{6}{|c|}{$U_c(p,H)$}\\
\hline
$H$ & $p=1$ & $p=0.9$ & $p=0.8$ & $p=0.75$ & NoSha \cite{cali2018percolation} \\
\hline
4.44 & 16.23 & 17.39 & 21.17  & 26.09 & 15.87 \\
2.67 & 7.07  & 8.30 & 10.59 & 13.72 &  7.44 \\
1.33 & 1.82 & 2.42 & 3.56 &  4.93 &  -- \\
0.89 & 0.41  & 0.77 & 1.48 & 2.41 &  1  \\
0.67 & 0  & 0.03 & 0.51 &  1.17 &  -- \\
0.53 & 0  & 0 & 0 & 0.45 & 0.32 \\
0.38 & 0 & 0 & 0 & 0 & 0.16 \\
\hline
\end{tabular}
\label{tab-critical-U}
\end{center}
\vspace{-4ex}
\end{table}

\indent Fig.~\ref{critical-user-estimation-twovar}(a) shows the variation of the critical user density $U_c(p,H)$ as a function of $H$ for several values of $p$. It is clear from Fig.~\ref{critical-user-estimation-twovar}(a) and Table \ref{tab-critical-U} that $H \mapsto U_c(p,H)$ is increasing for fixed $p$ and that $p \mapsto U_c(p,H)$ is decreasing for fixed $H$, which confirms that for given $H$, we can invert $U_c(p,H)$ to get back $p_c(U,H)$.

\section{Conclusion}
\label{s.Conclusions}
We have proposed a percolation model allowing one to study the connectivity 
of D2D  networks  in an urban canyon environment. It is
based on a  Poisson-Voronoi model of streets with canyon shadowing.
Poisson users on the edges (streets) and Bernoulli relays 
on the vertices (crossroads) establish  line-of-sight communications of bounded range on the streets.

This model allowed us to observe and quantify the following phenomena:
there is a minimal fraction of crossroads 
to be equipped with relays. Below this proportion, good connectivity of the network (indicated by percolation) cannot be achieved. Moreover, if the mean street length is not too big with  
respect to the communication range, 
then a small density  of  users  can  be  compensated by equipping more  crossroads with relays. 
If not, then good connectivity requires some minimal density of users compensated by the relays
in a way  explicitly estimated using our model.

While the precise critical values and functions 
certainly depend on the model, the general qualitative results 
(existence of the aforementioned regimes) are of more general nature and bring interesting  arguments to the discussion on the  possible D2D deployment scenarios.
In this regard, our work complements~\cite{cali2018percolation}, which 
does  not take into account any shadowing effects and thus does not predict
the strategic necessity of investments into relays located at crossroads to ensure connectivity between adjacent streets.
Concerning this necessary investment, observe that
the theoretical value of at least $71,3\%$ equipped crossroads might be 
smaller in practice. Indeed, in our model, crossroads are punctual. In reality, they have a certain surface and one well-placed regular user could ensure connectivity between adjacent streets.
Taking this into account would improve our quantitative predictions and is a track to follow for future work.

Other natural model extensions include more general shadowing,
e.g. via introducing two D2D connectivity radii, one for LOS connections, the other for non-line-of-sight (NLOS) connections. Other street system models, such as Poisson-Delaunay tessellations (PDT) or Manhattan grids (MG) could also be considered \cite{gloaguen2006fitting}.
Finally, introducing interference effects and user mobility in our model would definitely lead to more realistic predictions.  

\input{cameraready.bbl}

\end{document}